\documentstyle[11pt,aasms4]{article}

\lefthead{Lawson et al.}
\righthead{The fall and rise of V854 Centauri}

\begin{document}

\title{The fall and rise of V854 Centauri:\\ long-term ultraviolet
spectroscopy of a highly-active\\ R Coronae Borealis star}

\author{Warrick A. Lawson and Marco M. Maldoni}
\affil{School of Physics, University College, Australian Defence
       Force Academy,\\ Canberra ACT 2600, Australia;\\
       wal, mmmm@ph.adfa.edu.au}

\author{Geoffrey C. Clayton and Lynne Valencic}
\affil{Department of Physics and Astronomy, Louisiana State 
       University,\\ Baton Rouge, LA 70803;\\
       gclayton, valencic@rouge.phys.lsu.edu}

\author{Albert F. Jones}
\affil{Carter Observatory, P. O. Box 2903, Wellington, New Zealand;\\
       afjones@voyager.co.nz}

\and 

\author{David Kilkenny, Francois van Wyk, Greg Roberts and Fred Marang}
\affil{South African Astronomical Observatory, P. O. Box 9,
       Observatory 7935, South Africa;\\
       dmk,--, grr, --@saao.ac.za}

\begin{abstract}
We examine long-term low-dispersion {\it IUE\,} SWP and LWP 
spectroscopy of the R Coronae Borealis (RCB) star V854 Cen, 
obtained across the deep ($\Delta V > 6$ magnitudes) 1991, 
1992--1993 and 1994 declines.  We also report the optical 
light curve for the star in the interval 1987--1998, including 
multi-color photometry obtained during 1989--1998.  The light
curve includes at least 8 major declines where the amplitude 
exceeds 5 magnitudes, many of which appear to be multiple 
decline events. 

Analysis of the UV emission line spectra indicates most lines 
decay during the deep declines on characteristic timescales 
comparable to that reported for optical features.  Fe, Mg and
neutral C lines decay on timescales of typically 50--100 d.
Other lines, notably ionized C lines, decay on longer timescales 
($>$ 200 d) or appear to be unaffected by the declines.  The 
general nature of the UV emission lines and other UV features 
during the declines is consistent with the E1/E2/BL line-region 
model developed from the behavior of optical spectral features 
during declines.  However, the detailed line-behavior indicates
large intrinsic variability between decline events inconsistent
with the simple E1/E2/BL model.  Limited temporal coverage 
prevents detailed examination of the geometry of the emission 
line region or the obscuring dust.  We also report the first 
detection of the transition-region line \ion{C}{4}] $\lambda$1550 
in the spectrum of an RCB star.

We fit the onset times of all declines from maximum light within 
the 1987--1998 interval, irrespective of decline amplitude, with
a 43.23-d linear solution, thus improving the decline ephemeris 
of Lawson et al. (1992, MNRAS, 256, 347).  The linear term is 
probably the pulsation period of V854 Cen, further supporting 
the suspected link between radial pulsations and mass loss in 
these types of stars.

\end{abstract}


\section{Introduction}

The UV spectrum of an RCB star at maximum light closely resembles the
UV spectra of late F-type supergiants.  The most prominent features 
of the spectra are absorption lines of \ion{Mg}{2} $\lambda$2800 and 
\ion{Fe}{2} multiplets around 2400 and 2600 \AA.  The low resolution 
of most of the {\it International Ultraviolet Explorer\,} ({\it IUE\,}) 
spectra make line identifications difficult, particularly in 
the early-decline spectrum when many blended emission lines are present.  
Evans et al. (1985) attempted to identify the UV lines in an RY Sgr 
decline spectrum.  Clayton et al. (1992a) identified many lines in 
the 1991 decline of V854 Cen.  Holm et al. (1987) noted the similarity 
between the solar chromospheric spectrum and emission seen in the 
1983 decline of R CrB.  Table 3 of Clayton et al. (1992b) contains a 
line list prepared by comparing the decline emission spectra of RY Sgr, 
R CrB and V854 Cen.

Clayton et al. (1992b) summarized all previous spectroscopic work in 
the UV and visible on RCB stars, covering 10 declines of 3 RCB stars; 
R CrB, RY Sgr and V854 Cen.  Most studies have involved observations 
of a small portion of a decline.  In the visible spectral region, 
only the 1967 decline of RY Sgr (Alexander et al. 1972) and the 1988 
decline of R CrB (Cottrell et al. 1990) had good coverage from early 
in the decline until the return to maximum light.  Prior to the 
{\it IUE\,} observations reported in this paper, only the 1982--1983 
and 1990--1991 declines of RY Sgr had reasonable coverage in the UV   
(Clayton et al. 1992b). 

The general behavior of the emission spectrum in RCB declines is well 
known.  As the photospheric light is extinguished by the forming dust 
cloud, a rich narrow-line emission spectrum appears.  In the visible, 
this spectrum consists of many lines of neutral and singly ionized 
metals including \ion{Mg}{1}, \ion{Si}{1}, \ion{Ca}{1}, \ion{Sc}{2}, 
\ion{Ti}{2}, \ion{V}{2}, \ion{Cr}{2}, Mn I, \ion{Fe}{1}, \ion{Fe}{2}, 
\ion{Sr}{2}, \ion{Y}{2}, \ion{La}{2}, \ion{Ba}{2}, and \ion{La}{2} 
(Alexander et al. 1972; Payne-Gaposchkin 1963).  Most of the lines in 
this spectrum, referred to as E1 (Alexander et al. 1972), are short-lived.   
Within a few days or weeks, most of these lines have faded and are 
replaced by a simpler broad-line (100--200 km\,s$^{-1}$) spectrum 
dominated by \ion{Ca}{2} H and K, and \ion{Na}{1} D.  Some of the 
early-decline spectral lines remain strong for an extended period of 
time.   These lines, also narrow and referred to as E2, are primarily 
multiplets of \ion{Sc}{2} and \ion{Ti}{2}.  In particular, the 
\ion{Sc}{2} (7) $\lambda$4246 line remains strong.  The E2 lines 
are primarily low excitation.  There are many \ion{C}{1} absorption 
lines which fill-in but never go into emission (Alexander et al. 1972). 
The Balmer lines, which are typically very weak due to the hydrogen 
deficiency in these stars, do not go into emission.  The exception 
is in V854 Cen, which is much less hydrogen-deficient than the other
RCB stars.   V854 Cen shows strong Balmer line absorption at maximum 
light and emission in decline (Kilkenny \& Marang 1989; Whitney et al.
1992).  The late-decline spectrum is dominated by 5 strong broad-lines, 
\ion{Ca}{2} H and K, the \ion{Na}{1} D lines and a line at 3888 \AA\,
that may be He I.   The broad-line (BL) emission spectrum remains 
visible until the star begins to return to maximum light and the 
photospheric continuum regains dominance.

The UV spectrum undergoes a very similar evolution to that in the 
optical (Clayton et al. 1992b).   However, because all of the {\it IUE\,} 
observations during declines were made at low resolution, there is 
no information on the width of the emission lines.  The UV spectral 
evolution is most clearly seen in the 1991 decline of V854 Cen 
(Clayton et al. 1992a) but can also be seen in the other UV declines.  
The very early-decline UV spectrum consists of blends of many 
emission lines, primarily multiplets of \ion{Fe}{2} which make up a 
pseudo-continuum.  The \ion{Mg}{2} $\lambda$2800 doublet is present but 
not yet strong.  The strong apparent absorption at 2650 $\rm \AA$ 
is probably an absence of emission similar to that seen in the solar 
chromosphere (Holm et al. 1987).   The actual photospheric continuum 
is the bottom of this apparent absorption feature.   The V854 Cen 
spectrum is characterized by strong \ion{C}{2}] $\lambda$2325 emission 
(Clayton et al. 1992a).  With time, the early-decline spectrum begins 
to fade and be replaced by the late-decline spectrum.  In the transition 
between these two spectral phases, there is still much blended 
emission but \ion{Mg}{2} $\lambda$2800, \ion{Mg}{1} $\lambda$2852 and 
some of the \ion{Fe}{2} lines have started to become relatively stronger.  
The late-decline spectrum is characterized by blended emission from 
multiplets of \ion{Fe}{2} (2) $\lambda$2400, \ion{Fe}{2} (1) $\lambda$2600, 
\ion{Fe}{2} (62, 63) $\lambda$2750, as well as from \ion{Mg}{2} and 
\ion{Mg}{1}.  In addition, V854 Cen shows \ion{C}{1} and  \ion{C}{2}] 
emission which is generally not seen in the other stars.  Emission at 
\ion{C}{2} $\lambda$1335 is visible at maximum light in R CrB and 
RY Sgr (Holm et al. 1987; Holm \& Wu 1982).  Rao, Nandy \& Bappu (1981) 
report that emission is visible in \ion{Mg}{2} in R CrB at maximum
light.  

The data presented here represent the most extensive coverage 
of an RCB star ever obtained in the UV during declines.  54 LWP 
low-resolution, 2 LWP high-resolution, and 13 SWP low-resolution 
spectra were obtained from the archive.  Nearly half of these 
spectra were obtained when V854 Cen was 3 or more magnitudes below 
maximum light.  

\section{Observations and data reduction}

SWP and LWP spectra of V854 Cen were obtained with the {\it IUE\,}
satellite during 1991--1994.   These are listed in Table 1, along 
with the FES and estimated visual magnitude of the star at the 
time of the {\it IUE\,} observation.  Most of these data are 
large-aperture LWP and SWP low-resolution spectra.  All these 
files have been reprocessed with NEWSIPS.  Clayton et al. (1993) 
reported large-scale changes in UV line profiles in V854 Cen that 
appeared to be phased to the 43.2-d period of the star (Lawson 
et al. 1992).  The NEWSIPS (and INES; Schartel \& Skillen 1998) 
reduction of the {\it IUE\,} spectra shows no such effects.  The 
line profile changes observed were likely an artifact of the 
previous reduction due to the low signal-to-noise, and nature 
(the spectra consist of weak emission lines with weak or
absent stellar continuum) of these data.

The light curve of V854 Cen covering the interval 1987--1998, 
starting from near the discovery of the RCB-nature of the star 
(McNaught \& Dawes 1986), is shown in Fig. 1.  The star has been 
almost continuously monitored visually by one of us (AFJ) since 
discovery.  Fig. 1 shows over 1600 visual estimates made with a 
0.3-m telescope; the estimates obtained on average every 2.5~d.  
Sterken \& Jones (1997) discuss the visual observing procedure.
The uncertainty in the visual estimates is, at best, 0.1 mag,
but can rise to 0.3--0.5 mag when the star is varying rapidly 
in brightness or color.  Also, the visual estimates have an 
effective faint limit of $V \approx 13.5$.

Table 2 lists 323 sets of {\it UBVR$_{c}$I$_{c}$} photometry 
of V854 Cen obtained with the 0.5-m telescope at the South African 
Astronomical Observatory during 1989--1998.  The measurements
of V854 Cen were tied to observations of E-region photometric 
standards, and most of the $V$ magnitudes and colors have 
1-$\sigma$ uncertainties of $<$ 0.01 mag.  Measurements of 
lower quality (normally when the star is faint) are given to 
correspondingly lower precision in Table 2.  The $V$ light 
curve is shown in Fig. 1 and is discussed in Section 3.1; 
the color curves are shown in Fig. 2 and are discussed in 
Section 3.2.

Figs 1 and 2 also show decline onset-times used to improve the 
Lawson et al. (1992) decline ephemeris of V854 Cen, which we 
discuss in detail in Section 4.  Fig. 3 shows the 1991--1994 
photometry and visual estimates on an expanded scale.  Note
the generally good correspondence between the photoelectric
and visual data, where these overlap.  Times of the {\it IUE\,} 
LWP and SWP observations are indicated in both Figs 1 and 3.   

\section{Description of the observations}

\subsection{Linking the light curve and the {\it IUE\,} spectra}

The discovery decline (1987; commencing near JD6970; where Julian
Dates are given as the difference JD--2440000) of V854 Centauri 
(then known as NSV 6708) was only observed visually.  The 1988 
decline (JD7400) was observed by Kilkenny \& Marang (1989) and 
Lawson \& Cottrell (1989), who obtained {\it UBVRIJHKL\,} 
photometry and visual spectroscopy.  The 1989--1991 light curve 
has been discussed in detail by Lawson et al. (1992).  The 1991 
decline appeared to consist of three separate events (near JD8310, 
8350 and 8395) with respective minima of $V \approx$ 11, 14 and 15.  
The first series of {\it IUE\,} spectra (1991; JD8335--8500) were 
acquired during the rise from the first fade, soon after the second 
minimum and throughout the deep third minimum.  The final 1991 
{\it IUE\,} spectra were made at $V \approx$ 10.0 as the star 
began to recover towards maximum light.

The star recovered to maximum light ($V$ = 7.3) only briefly
near JD8610, before fading to $V$ = 13.5 by JD8750.  Structure in
the light curve is apparent during the initial decline (Fig. 3),
with two partial recoveries near JD8700 and JD8730, then again at
minimum near JD8810.  The 1992 series of {\it IUE\,} spectra follow the 
decline from JD8650 ($V$ = 7.7) through JD8868 ($V \approx$ 13).

The star experienced a prolonged minimum until JD8980 (1993 January).
There are no photometric measurements from JD8850 until JD9030 by
which time the star was $V$ = 9.0.  The few visual estimates made
between JD8850--8980 indicate the star was highly-variable, and to 
have briefly reached $V \approx$ 10 near JD8960.  There were no UV 
spectra during this period.  

The star regained maximum light near JD9100.  The rise to maximum, 
and the time at maximum light, was well-monitored in the UV; the 
1993 series of {\it IUE\,} spectra consist of 27 observations made 
between JD9009--9217.  High-resolution LWP spectra were obtained 
on JD9093 and JD9140 when V854 Cen was at, or near, maximum light 
(Lawson et al., in preparation).  1993 was the only year in the data 
set not characterised by the onset of a deep decline; however the star 
experienced a relatively long ($\sim$ 200 d) low-amplitude decline 
from JD9180--9350 that was unique in the 12-yr light curve.  

The early-1994 light curve was characterised by the steepest decline 
in our data set.  The $V$ magnitude decreased from 7.5 (at JD9430)
to 13 in $\sim$ 20 d.   11 {\it IUE\,} spectra were obtained during this 
time; from JD9427--9456.  V854 Cen remained at minimum for $\sim$
30 d before rising to $V$ = 9 near JD9530.  The star subsequently
faded again; slowly to $V$ = 10 near JD9600, then rapidly to
$V$ = 13.5 near JD9620.  The final 6 {\it IUE\,} spectra were obtained
from JD9508--9539.

The remainder of the light curve was characterised by a major
decline commencing in 1994 November (JD9860) with the star not
fully-regaining maximum light until mid-1997 (near JD10600).
Short-duration fades were seen near JD10345 and JD10775.  The
star entered a deep decline on JD10855 (1998 February).

\subsection{The color curves}

In Fig. 2 we present the color curves of V854 Cen from 1989--1998 
(JD7500--11100). The colors during 1989--91 have been discussed by 
Lawson et al. (1992).  The longer-term trends in the colors are 
similar to that already reported for this star and other RCB stars.  
For instance, in the 1990, 1991 and 1992 decline events, while the 
($U-B$) decreased, the other colors reddened and varied in sympathy 
with the $V$ curve (Figs 1 and 3). This effect is understood as the 
($U-B$) index being `driven' by the appearance of the emission line 
region which becomes prominent as the ejected dust cloud obscures 
the photosphere.  Cottrell et al. (1990) observed that the ($U-B$)
color turned blueward upon the appearance of the E1 emission  
spectrum during the 1988 decline of R CrB.  Sometimes both the 
($U-B$) and ($B-V$) colors decrease during the early stages of the 
decline before eventually reddening.  Cottrell et al. (1990) termed 
these events `blue' declines.  

The 1994 and 1998 declines showed significant variations in the 
($U-B$) color.  The former decline is characterized by all the 
colors displaying blueward trends which correlate with the extreme 
nature of the event, i.e., a rapid decrease in visual light to 
$V \sim$14 in $\sim$ 15~d.  Presumably the photosphere was 
obscured rapidly without significant obscuration of the emission
line region.  On JD9440, 10~d after the decline onset when the
star was $V$ = 13.4, the ($U-B$) color was 0.9 mag bluer (at --0.5)
than at maximum (0.4) and the ($B-V$) color was 0.5 mag bluer
(0.0 {\it cf.}\, 0.5 at maximum).  The next photometry was obtained 
at JD9485, by which time the star had recovered to $V \approx$ 12.

The color behavior during the 1998 minimum differed in that the 
($U-B$) and ($B-V$) colors become bluer during the decline minimum,
at a time when the ($V-R$) and ($V-I$) colors were rapidly reddening.
The ($U-B$) color peaked at --0.5 on JD10947, when the star was $V$
= 13.5; 92~d after the decline onset at JD10855.  Extreme color 
variations at minimum are probably due by the emergence of the 
optical BL spectrum (lines such as Ca H and K and Na D and broad 
continuum emission; see, e.g. Cottrell et al. 1990, figs. 2--4 and 
Clayton et al. 1993, fig. 2) as well as optical depth variations 
in the dust and coverage of the photosphere.  However, the decline 
is not a `blue' decline following the Cottrell et al. (1990) 
description.

Cottrell et al. (1990) also describe `red' declines where both 
the photospheric and chromospheric fluxes are simultaneously 
reduced.  This results in the color indices increasing, i.e. 
reddening.   The 1992 decline appeared to show this trend.  The
initial behavior of the colors during the 1998 decline was also
redward.

\subsection{The UV spectrum evolution} 

A description of the UV spectrum of V854 Cen at maximum and minimum 
light was given by Clayton et al. (1992b). Lines of interest in the 
SWP and LWP spectra include \ion{C}{2} $\lambda$1335, \ion{C}{3}] 
$\lambda$1909 and \ion{C}{2}] $\lambda$2325, \ion{Fe}{2} multiplets 
at $\lambda$2400, 2600 and 2750, \ion{Mg}{2} $\lambda$2800, \ion{Mg}{1} 
$\lambda$2852 and \ion{C}{1} $\lambda$2965.  Some of these are 
known to vary in strength as V854 Cen goes into a decline.  Gross 
variations in the appearance of the SWP and LWP spectra are shown in 
Figs 4 and 5, where we show these spectral regions at maximum 
and minimum light.  The NEWSIPS reduction has revealed the \ion{C}{4}]
$\lambda$1550 transition-region line for the first time in the spectrum
of V854 Cen, and the only detection of this line in an RCB star.  
{\it IUE\,} spectra of RY Sgr and R CrB of similar signal-to-noise 
show no feature at this wavelength.   Clayton et al. (1999) observed 
RY Sgr with STIS in the far-UV ($\lambda\lambda$1140--1740 \AA\,) 
and observed strong \ion{C}{2} $\lambda$1335 and \ion{Cl}{1} 
$\lambda$1351 emission, and possibly fluoresced CO emission pumped 
by \ion{C}{2} $\lambda$1335.  There was no indication of \ion{C}{4}] 
$\lambda$1550 in the STIS spectrum of RY Sgr.

The acquisition of UV data over an extended time span such as that 
covered by these observations may uncover long-term trends in the UV 
spectral evolution, particularly the quantitative changes in the 
strengths of the lines during declines.  The long series of UV decline 
spectra of V854 Cen is unique for an RCB star, and is unlikely to be 
surpassed by current platforms such as {\it HST\,}.

Only a few measurements of ultraviolet emission-line strengths have 
been published.  Herbig (1949) reported that the strength of the 
\ion{Ca}{2} H and K lines of R CrB peaked as the star went into a
decline and subsequently faded to one-fifth of their original strength. 
In contrast, the \ion{C}{2} $\lambda$1335 line in the same object was 
found to roughly remain constant during the 1983 decline at the same 
value measured at maximum light (Holm et al. 1987). Clayton et al. 
(1992b) provided further data concerning the strength of the \ion{Mg}{2} 
emission line for the 1983 and 1988 decline of R CrB, the 1982 and 1990 
decline of RY Sgr, and the early 1991 decline of V854 Cen. They found 
in R CrB that the line peak flux remained constant for 200~d into 
a decline, while in RY Sgr and V854 Cen the emission strength appeared 
to decrease after $\sim$ 100~d in a manner similar to that observed 
by Herbig (1949) for the \ion{Ca}{2} H and K lines in R CrB.  

We have measured the peak line strengths in the {\it IUE\,} spectra 
of V854 Cen.  Over the course of these observations (1991--1994) 
the \ion{C}{2} $\lambda$1335, \ion{C}{4}] $\lambda$1550, \ion{C}{3}] 
$\lambda$1909 and \ion{C}{2}] $\lambda$2325 have average strengths 
of $1\times10^{-14}$, $7\times10^{-15}$,  $1\times10^{-14}$ and 
$4\times10^{-14}$, respectively, as seen at the {\it IUE\,} 
resolution (units are erg\,s$^{-1}$\,cm$^{-2}$\,\AA$^{-1}$).  The 
\ion{Fe}{2} $\lambda$2600, 2750 lines were at their maximum strengths 
during 1991 ($8\times10^{-15}$ and $5\times10^{-15}$, respectively). 
Subsequently, throughout the data set, they were generally weaker 
than these values. The \ion{Mg}{1} and \ion{Mg}{2} lines, which were 
generally present during all minima, had respective maximum fluxes 
in 1991 of $1.5\times10^{-14}$ and $5\times10^{-15}$.  Finally, 
\ion{C}{1} $\lambda$2965 had a similar maximum peak flux during 
1991, 1992 and 1994 of $4\times10^{-15}$.

We have also measured the integrated flux (line intensities) of 9 
key emission features in the SWP and LWP spectra except where the 
rising background continuum (at or near maximum light in the SWP 
spectra, and always present to some extent in the LWP spectra) made
continuum-subtraction measurements unreliable.  The line intensities
of the \ion{C}{2} $\lambda$1335, \ion{C}{4}] $\lambda$1550 and 
\ion{C}{3}] $\lambda$1909 lines (in the SWP spectra), and the 
\ion{C}{2}] $\lambda$2325, \ion{Fe}{2} $\lambda$2400, 2600, 2750, 
\ion{Mg}{2} $\lambda$2800, \ion{Mg}{1} $\lambda$2852 and 
\ion{C}{1} $\lambda$2965 lines (in the LWP spectra) are shown in 
Fig. 3 (units are erg\,s$^{-1}$\,cm$^{-2}$).  The uncertainty of a
typical measurement is 10--20 percent; the uncertainty is contributed 
to by the choice of continuum placement and the quality of the 
flux calibration.  Importantly, the trends seen in Fig. 3 agree with 
the visual assessment of the temporal evolution of the spectral 
features.  Only large-aperture LWP spectra are measured because of 
the uncertain photometric corrections needed for the small-aperture 
observations.  We have compared a number of SWP and LWP spectra
extracted using both NEWSIPS and INES (Schartel \& Skillen 1998)
and find minimal differences in the appearance of the spectra and 
the emission line fluxes between the two reductions.  

All the measured features are blends, and the intensities
plotted in Fig. 3 represent the integrated flux of the blend, e.g.
\ion{Fe}{2} $\lambda$2600 consists of several weak \ion{Fe}{2} lines 
with wavelengths ranging from 2586 to 2631 \AA\, (see Fig. 5; some of 
these lines are resolved); \ion{Mg}{2} $\lambda$2800 consists of a 
doublet at $\lambda$2796, 2803.  Wu et al. (1992; see table 1.1) 
identifies many of these blends.  Clayton (1992b;
see table 3) list the presence of these lines across a number of
declines for several RCB stars.  There is no indication in the 
{\it IUE\,} spectra that the relative contribution of lines composing 
the blends changes as the blend intensity varies with time. 

\subsubsection{The 1991 decline}

The 1991 decline (from JD8310) is characterised by 3 progressively
deeper fades with the first {\it IUE\,} spectrum obtained near the 
first local minimum.  All the LWP emission lines generally weakened 
during the interval JD8335--8500 (see Fig. 3).  Line-intensity 
half-lives range from $\sim$ 50 d for \ion{Fe}{2} $\lambda$2750, 
\ion{Mg}{2} $\lambda$2800 to $\sim$ 200 d for \ion{C}{2}] $\lambda$2325
(see Fig. 6 and Table 3).  Several features reached minimum strength 
at JD8436, in agreement with the time of the light curve minimum 
(Fig. 3).  In particular, \ion{Fe}{2} $\lambda$2600 increased in 
strength by a factor of $\sim$ 4 from the light of minimum light 
($V \approx$ 15) as the star brightened to $V$ = 10.  Lines such 
as \ion{Fe}{2} $\lambda$2750 continued to decay during this time.  
In contrast to general behavior of other LWP spectral lines, the 
\ion{Mg}{1} $\lambda$2852 may have strengthened slightly as the star 
passed through the first 2 secondary minima and after the onset of 
the third minimum at JD8395.  Thereafter, as the star faded to 
$V = 15$, the line intensity also rapidly decreased. In the 1991 
SWP spectra, obtained after the decline minimum, all three key
lines (\ion{C}{2} $\lambda$1335, \ion{C}{4}] $\lambda$1550 and 
\ion{C}{3}] $\lambda$1909) appeared to remain constant.  

\subsubsection{The 1992--1993 decline}

The 1992--1993 decline was the longest in our data set, lasting 
$\sim$ 400 d.  Similar to many other declines of V854 Cen, this 
event shows multiple (in this case, 2) local minima between the 
onset of the decline from maximum light ($V$ = 7.3) at JD8610 and
minimum ($V \approx$ 14) near JD8750.  SWP and LWP spectra were 
obtained during the fade; other LWP spectra were obtained near 
minimum when the star brightened briefly to $V$ = 12 near JD8810.
SWP and LWP spectra were obtained during the rise to maximum
light during 1993, including 2 LWP high-resolution spectra.

The \ion{C}{2} $\lambda$1335 line decayed by a factor of $\sim$ 
3 during the fade.  The behavior of the line is unusual compared 
to the 1991 observations which showed the line to be bright 
(comparable to the intensity of the line at maximum light) even 
though the observations were made when the star was fainter than 
at any time during the 1992 decline.  The next SWP observations 
were not made until JD9009, just prior to the star beginning to 
brighten towards maximum light; thus no SWP spectra were obtained 
across the 200 d minimum.  During the rise, the \ion{C}{2} 
$\lambda$1335 line intensity was slightly ($\sim$ 50 percent) 
greater than when measured at the decline minimum.  However, 
it remained a factor of $\sim$ 2 weaker than during 1991. 
\ion{C}{4}] $\lambda$1550 may have been fainter during the 
decline minimum, based upon a single faint flux measurement 
at JD8787.  \ion{C}{3}] $\lambda$1909 was insensitive to the 
decline and it remained at a level similar to that seen in 1991.  

The LWP spectrum lines showed a variety of responses to the 
decline.  \ion{C}{2}] $\lambda$2325 remained relatively constant 
and possibly stronger than in 1991.  Fe and Mg lines were weak, 
but some brightened in sympathy with the 2 magnitude increase 
in visual flux near JD8820.  (The actual rise in flux may have 
been somewhat greater than 2 magnitudes since the decline minimum 
was not well-observed photometrically and the visual estimates 
are unreliable below $V \approx$ 13.5.)  \ion{C}{1} $\lambda$2965 
showed somewhat different behavior to that of the other lines.  
Measurements made during the final fade to minimum light in 1992 
showed the line strength decreased from $\sim 5\times10^{-14}$ 
erg\,s$^{-1}$cm$^{-2}$ to unmeasureably low ($< 5\times10^{-15}$) 
between JD8720 and JD8724, and then regained its former strength 
by JD8792.  At the time of the local maximum near JD8820, the 
line appeared to decrease in strength by 30--50 percent before 
recovering.  

During the rise to maximum light during early-1993 (JD9000--) most
LWP spectra are affected by the rising stellar continuum and the line 
strengths are unreliable or unmeasureable.  However, on JD9009, most 
LWP lines had brightened above the average level measured near JD8800.

\subsubsection{The 1994 decline}

The 1994 decline was the most extreme of our data set, taking only 
$\sim$ 16~d for the star to fade from maximum light ($V \approx$ 
7.5) at JD9430 to minimum ($V \approx$ 13.50) at JD9446.  As in the 
previous decline, the absence of photometric data through the decline
minimum did not allow an accurate determination of the amplitude 
of the event.  Visual estimates indicated the star remained
at or below $V$ = 13.5 for $\sim$ 30~d.  Subsequently, the star
partly recovered (to $V$ = 9, including a minor fade), which was
then followed by another deep decline.  Most of the LWP spectra 
acquired during 1994 were taken during the onset of the first 
decline with 3 LWP and 2 SWP spectra monitoring the recovery phase 
of the light curve.  The SWP spectra showed that \ion{C}{4}] 
$\lambda$1550 and \ion{C}{3}] $\lambda$1909 remained at a strength 
comparable to that during the 1991 and 1992--93 declines.  However, 
the strength of \ion{C}{2} $\lambda$1335 was similar to that during 
the 1992--93 decline. Presumably all the LWP line strengths decreased 
rapidly as the optical flux of V854 Cen faded towards minimum.  
This behavior was only observed in \ion{C}{2}] $\lambda$2325.  
Measurement of the line at JD9437 (7~d after the decline onset) 
gave an intensity of $5\times10^{-13}$ erg\,s$^{-1}$\,cm$^{-2}$.  
By JD9449 (19~d after the decline onset) the line strength was 
comparable to the level recorded during the 1991 decline minimum 
($2.5\times10^{-13}$ erg\,s$^{-1}\,$cm$^{-2}$).  By JD9449, all 
other LWP spectral lines were weak.  With the possible exception 
of \ion{Fe}{2} $\lambda$2600, all lines remained weak during the final
series of {\it IUE\,} observations obtained as the star rose in 
brightness to $V$ = 9 by JD9540.

\subsection{Summary of line behavior}

Trends in the line strengths presented in Fig. 3 suggest that it may
be possible to group some spectral lines according to their responses
to a decline event.  Clayton et al. (1992b) attempted to relate changes 
in the UV spectrum to those seen in the visible spectrum (Alexander 
et al. 1972, Cottrell et al. 1990).  In their scheme, the ultraviolet
E1, E2 and BL spectral features roughly correspond to their visible 
counterparts and were defined as:

\noindent
E1 (fade in 10--30~d): Many blended lines of \ion{Fe}{2} and other 
ionized metals, the apparent absorption feature at 2650 \AA\, and 
weak \ion{Mg}{2} $\lambda$2800.

\noindent
E2 (fade in 50--150~d): Most of the E1 lines have faded leaving 
\ion{Fe}{2} multiplets at $\lambda$2400, 2600 and 2750.  

\noindent
BL (fade but never disappear): Long-lasting lines of \ion{C}{2}] 
$\lambda$2325, \ion{Mg}{2} $\lambda$2800, \ion{Mg}{1} $\lambda$2852 
and \ion{C}{1} $\lambda$2965.

In the 1991 decline, during the JD8300--8500 interval, \ion{C}{2} 
$\lambda$1335, \ion{C}{4}] $\lambda$1550 and \ion{C}{3}] $\lambda$1909 
are clearly BL as they show little change in strength during the 
decline minimum.  \ion{C}{2}] $\lambda$2325 also probably originates 
in the BL emission region although it does weaken by $\sim$ 50 
percent (Fig. 6).  

The \ion{Fe}{2} $\lambda$2750, \ion{Mg}{2} $\lambda$2800, 
\ion{Mg}{1} $\lambda$2852 and \ion{C}{1} $\lambda$2965 lines all 
undergo significant decreases in intensity over the 170 day {\it IUE\,} 
coverage and thus are classified as E2 (Fig. 6).  \ion{Mg}{1} 
$\lambda$2852 may be stronger near JD8400, in phase with the local 
maximum (Fig. 3).  This may indicate that its region of origin is 
closer to the star than is the case for the other E2 lines.  The 
behavior of \ion{Fe}{2} $\lambda$2600 is unusual as, unlike other 
E2 lines, it recovers strongly after only $\sim$ 100~d (Fig. 3).  
However, the decrease in intensity seen between JD8360--8466 is 
not rapid enough to classify the line as E1 and hence it is 
considered to be E2 (Table 3).  

The only time an E1 LWP spectrum was obtained during 1991 was at 
JD8335.  This spectrum looks similar to the early-decline spectra 
of RY Sgr and R CrB (Clayton et al. 1992a) as it showed a myriad 
of blended \ion{Fe}{2} lines, apparent absorption at 2650 \AA\, 
and \ion{Mg}{2} $\lambda$2800 substantially filled-in by emission.  
The first two E1 features were absent in the next LWP spectrum 
(JD8361) with \ion{Mg}{2} $\lambda$2800 being stronger (in emission) 
than at any other time in the data set.   Clayton et al. (1992b) 
noted when there are several local minima, the E1 spectrum does 
not generally reappear, unless there is a long time between the 
local minima.   During the 1991 decline, E1 features are only seen 
during the first local minimum.  No {\it IUE\,} spectra were obtained 
between JD8500 (after the end of the 1991 minimum) and JD8650 (near 
the onset of the 1992--93 decline).  During this time, the spectrum 
showed full recovery from one dominated by weak emission to one 
swamped by deep photospheric absorption.  (Figs 4 and 5 show the 
gross changes in the SWP and LWP spectra from that at maximum 
light to the spectrum during the decline minimum.)   
\ion{Mg}{2} $\lambda$2800 best demonstrates 
the transition (Fig. 5).  Visual inspection of the spectra (where 
line strengths could not be reliably measured) further supports the 
E2-type nature of this line. 

The behavior of some of the lines during the 1992-93 decline shows 
the difficulty in trying to uniquely characterize their nature using 
emission-line decay times, e.g. the correlation between the light 
curve and the timescale for the decrease in \ion{C}{2} $\lambda$1335 
is classic E2 (Table 3), unlike the BL appearance of this line in 
1991.  Inspection of Fig. 3 confirms the BL nature of \ion{C}{2}] 
$\lambda$2325 and \ion{C}{3}] $\lambda$1909; \ion{C}{4}] $\lambda$1550 
is probably BL.  The timescales and appearance of the Mg lines suggest 
that these are E2 lines. Inspection of the LWP spectra shows that, 
during the multiple fade to minimum, \ion{Mg}{2} $\lambda$2800 emission 
progressively fills-in the deep \ion{Mg}{2} $\lambda$2800 absorption 
feature, but remains weak or absent and only becomes prominent during 
the decline minimum.  During the rise to maximum light in 1993 it 
appeared as a weak emission feature on JD9009, whereas it was visible 
(in emission) at the bottom of a weak photospheric absorption at 
JD9045.  Unlike the 1991 decline, the 2650 \AA\, absorption feature  
was absent during this event. 

The behavior of the \ion{Fe}{2} $\lambda$2600, 2750 and 
\ion{C}{1} $\lambda$2965 lines (Fig. 3) was more difficult to interpret. 
In particular, it was difficult to reconcile their changes in strength 
(and timescales) assuming the E1/E2/BL model.  The rapid decay of 
\ion{C}{1} $\lambda$2965 at JD8720 is compatible with E1-type behavior.  
No other LWP lines showed this rapid response to the decline minimum.  
These lines show rapid and discordant behavior during the local maximum 
near JD8820.  \ion{Fe}{2} $\lambda$2720 appeared to peak at JD8829; 
whereas \ion{C}{1} $\lambda$2965 decreased in strength by a factor of 
$\sim$ 2 near JD8829/8835.  Both lines recover their pre-JD8829 fluxes 
by JD8868.  The timescales of both lines are more suggestive of E1 than 
E2.  \ion{Fe}{2} $\lambda$2600 showed little or no reaction to the local 
minimum.

Finally, measurements made during the 1994 decline presented a similar
classification dilemma to that discussed above.  In this event, 
all SWP lines remained at strengths similar to that observed during
previous declines, and thus are BL, but all LWP lines faded rapidly 
and apparently in phase with the light curve.  The decrease in the 
line intensity was only measured in CII] $\lambda$2325 (Fig. 3 and 
Section 3.3.3) on timescales (the line faded by a factor of 2 in 
20~d) consistent with E1 behavior, when the line either showed 
long-E2 or BL behavior during 1991-93.  By JD9449, 19~d after 
the decline onset, all LWP lines were faint at levels similar to 
that observed during the 1991 and 1992-93 decline minima (Fig. 3).
Thus we would conclude from the apparent timescales for the fading
of these lines that all these features are E1, if it were not for
the combined photometric and spectroscopic evidence that the 1994 
decline was particularly rapid compared to the other declines.  The
color curves (Fig. 2) indicate that the ($U-B$) color was 0.9 mag 
bluer than at maximum at JD9440, only 10~d after the decline 
onset -- evidence that the photosphere was rapidly obscured, 
exposing the emission line region.  The LWP spectrum obtained on 
JD9449 indicated that the emission line region was then obscured 
on a timescale of $<$ 9~d.  From inspection of the LWP spectra, 
we can also estimate that it took $<$ 19~d after the decline 
onset for the 2650 \AA\, feature to disappear and for the 
\ion{Mg}{2} $\lambda$2800 absorption line to weaken and fill-in, 
and then go into emission.

In summary -- while the emission line behavior across several
declines of V854 Cen was generally in agreement with the E1/E2/BL
model, the behavior is more complex than the simple model predicts.  
Temporal coverage of the declines is insufficient to discern the 
geometry of the emission line regions and that of the eclipsing 
dust, e.g. UV Mg lines were expected to behave like optical BL 
lines (such as \ion{Ca}{2} H and K, and \ion{Na}{1} D) but instead 
showed more-rapid activity commensurate with E1 and E2 lines.  As 
more data is obtained of these types of events, we expect these 
stars and their individual declines will probably show large 
intrinsic variation in the nature of the emitting region and 
the eclipsing dust cloud.

\section{A revised pulsation-decline ephemeris for V854 Cen}

Lawson et al. (1992) discovered a link between decline onset times
and the probable pulsation period of the star, with 8 decline onset
times between JD6970 (1987) and JD8395 (1991) being fitted by the
linear solution: 

\noindent
JD$_{n}$ = 2447400.6 ($\pm$ 1.1) + 43.2 ($\pm$ 0.1) $n$ d,

\noindent
between cycle numbers $n$ = --10 and 23. 

The ephemeris fitted the onset times of the 1987 (JD6970, $n = -10$) 
and 1988 (JD7400, $n = 0$) declines, two low-amplitude fades and one
large-amplitude decline in 1989 (JD7705, 7785 and 7875; $n$ = 7, 9 
and 11, respectively), and the triple 1991 decline (JD8310, 8350 
and 8395; $n$ = 21, 22 and 23, respectively).  The ephemeris also
appeared to satisfy times of maxima on the light curve during 1989
that may be due to low-amplitude pulsations 
of the star.  If this is the case, V854 Cen is similar to the RCB star 
RY Sgr, which has pulsation maxima and decline onset times tied to a 
37.8-d period (Pugach 1977).  A 43-day pulsation period for V854 Cen 
would be entirely consistent with the pulsation period of other RCB 
stars of similar $T_{\rm eff}$ most of which have pulsation periods 
of $\sim$ 40~d (Lawson et al. 1990, Lawson \& Cottrell 1997). 

With the light curve from 1987--1998 available to us (Fig. 1), we
have extended the Lawson et al. decline ephemeris across the entire 
data set.  We fit the onset times (15 epochs) of {\it all\,} declines 
from maximum light ($V \approx$ 7.5; 13 epochs) within the 1987--1998 
interval, irrespective of decline amplitude, and the last 2 fades of
the triple 1991 decline (the first fade occurred from maximum light
and is included in the 13 epochs above; the other two fades occurred 
from $V \approx$ 10), with the linear solution:

\noindent
JD$_{n}$ = 2447400.43 ($\pm$ 1.33) + 43.23 ($\pm$ 0.03) $n$~d,

\noindent
between cycle numbers $n$ = --10 and 80. 

The 15 epochs are indicated in Fig. 1, and some are shown in Figs 2 
and 3 where the epochs extend across the interval observed photometrically.  
The revised solution is in agreement with the Lawson et al. (1992) 
ephemeris to within the respective uncertainties of the two solutions.  
Table 4 lists the observed epochs, calculated epochs from the ephemeris, 
and the observed--calculated (O--C) residuals.  Figure 7 plots the O--C
residuals as a function of period number, and binned as a histogram.  
The 1-$\sigma$ scatter of all 15 residuals is 3.3~d.  The earlier 
($n$ = --10 to 28) residuals determined mainly from photoelectric 
measurements have lower scatter (1-$\sigma$ = 2.1~d) than the 
later residuals ($n$ = 41 to 80, 1-$\sigma$ = 4.9~d) which are 
determined mainly from the visual estimates.  However, all of 
these 1-$\sigma$ values are similar to the typical uncertainty 
in the observed decline onset time of $\pm$ 5~d.  There is 
currently no evidence for higher-order, e.g. quadratic, terms 
in the ephemeris, as has been claimed for RY Sgr (Kilkenny 
1982, Marraco \& Milesi 1982).  

Most of the declines showed complex structure in the light curve
as the star faded from maximim light, with several local maxima 
giving the appearance of multiple decline events.  Only in the 
1991 decline (commencing JD8310) did the times of the local maxima
seen during the fade (JD8350, 8395) support both the Lawson et al. 
(1992) and revised ephemerides.  Some of the structure seen during 
declines was approximately fitted by the revised ephemeris.  In a 
number of declines, local maxima seen during the initial fade, and 
during the decline minima, gave an O--C residual of --10 to --15~d,
i.e. the features had `maxima' that occurred 10--15~d before the 
ephemeris prediction.  Such features occurred during the 1992 (near 
JD8690, 8725, 8815), 1994 (JD9540, 9590) and 1998 (JD10885, 10930) 
declines.  The apparent connection between these times and the 
ephemeris, only offset by 10--15~d, suggests they are also linked
in some way to the 43.23~d periodicity.   

\section{Discussion}

\subsection{Dust formation in RCB stars}

The traditional model for the declines in RCB stars considers 
the homogenous nucleation of carbon particles in thermodynamic 
equilibrium at temperatures of $\sim$ 1500 K and at a distance of
$\sim 20 R_{\star}$ (see, e.g. Feast 1986).  Radiation pressure 
slowly dissipates the dust cloud.  

Over the past 10 years, a wealth of observational and theoretical 
evidence has pointed increasingly towards dust formation in the 
near-star field, at distances of $\sim 2 R_{\star}$.  For example, 
(i) simultaneous optical photometry and spectroscopy indicates 
the dust cloud can initially obscure only part of the photosphere.  
The cloud can then be rapidly accelerated away from the star by 
radiation pressure.  The photosphere is obscured on timescales of 
typically 10--20~d, revealing a rich emission line region of 
which the inner regions (the E1 spectral region) are obscured in 
a further 10--20~d (Cottrell et al. 1990, Lawson 1992).  The 
timescale of the fade to minimum light of 30--50~d, followed 
by recovery on timescales of hundreds of days, can be modelled by 
radial expansion at 100--200 km\,s$^{-1}$ of an obscuring cloud 
from a point of close proximity to the star.  The expansion velocity 
is consistent with that of blue-shifted absorption features seen 
during declines (Alexander et al. 1972, Cottrell et al.  1990, 
Clayton et al. 1993).  (ii) The decay timescales for lines formed 
in the emission region suggests the E1 region has an extent of 
1.5--2 $R_{\star}$, the E2 region is $\sim$ 5 times larger, and 
the BL region is larger still.  The available evidence suggests at 
least 2 temperature regimes in the emission line region; a cool 
($\sim$ 5000 K; Clayton et al. 1992a) inner region likely to be the 
site of neutral and singly-ionized species composing the E1 and E2
spectrum, and a much hotter outer region indicated by the presence 
of BL lines such as \ion{C}{3}] $\lambda$1909, \ion{C}{4}] 
$\lambda$1550 and \ion{He}{1} $\lambda$10830.  The presence of 
\ion{C}{4}] $\lambda$1550 implies the presence of a transition region 
with an electron temperature $T_{\rm e} \sim 10^{5}$ K (Jordan \& 
Linsky 1987).  
(iii) The linking of decline onset times and pulsation periods in 
RY Sgr (Pugach 1977) and V854 Cen (Section 4) is only possible if dust 
formation is intimately associated with the pulsations of the star, 
with little phase delay between the time of maximum light on the 
pulsation cycle and the onset of the decline.  Clayton et al. 
(1992b) reviewed empirical evidence for dust formation near the 
stellar surface.  

More recently, (iv) Woitke, Goeres \& Sedlmayr (1996) produced 
models that suggested the presence of (pulsation-induced) shocks 
in the outer atmosphere of a hydrogen-deficient star might result 
in conditions far-removed from thermodynamic equilibrium, encouraging 
particle nucleation.  Such photospheric shocks are observed in the
RCB star RY Sgr (Lawson, Cottrell \& Clark 1991, Clayton et al. 1994)
and may be present in other RCB stars.  (v) Lawson \& Cottrell (1997) 
showed all well-observed RCB stars were pulsating stars, and (vi) 
Clayton et al. (1999) reported the probable discovery in RY Sgr of 
CO, critical to the Woitke et al. (1996) model.  Polar molecules 
such as CO play a major role in gas radiative heating and cooling.  
In hydrogen-deficient atmospheres, CO is expected to be the most 
abundant polar molecule by two orders of magnitude.

To this evidence, measurement of UV emission lines in V854 Cen 
shows some consistency with the E1/E2/BL model developed from the 
behavior of the optical spectrum during declines.  Although we have 
poor sampling near the times of decline onset, the few E1 spectra 
obtained suggest characteristic E1 lines decay on timescales of 
several tens of days (see Section 3.4).  E2-region lines in 1991 
decay on timescales of 50--120~d (see Section 3.3.1).  BL-region 
lines throughout the data set generally decay on timescales of 
hundreds of days (e.g. \ion{C}{2} $\lambda$1335).  Some lines may 
be `super-BL' (coronal and transition-region lines such as 
\ion{C}{3}] $\lambda$1909 and \ion{C}{4}] $\lambda$1550) and 
remain essentially constant despite the high-degree of activity 
seen in the V854 Cen light curve.  

Uniquely for an RCB star, we have analysed spectroscopy across 
several consecutive major declines of V854 Cen.  Emission line decay 
timescales clearly differ between declines.  This can be related to 
probable free parameters such as the initial size and extent of the 
obscuring cloud, ejection velocity and future evolution of the cloud, 
and axis of the cloud motion with respect to the line-of-sight (e.g. 
Pugach 1990).  Hardly surprising,  the E1/E2/BL scheme needs to be 
interpreted as a simple classification scheme for characterising the 
evolution of the post-decline spectrum, where decay timescales and 
the behavior of individual lines should be seen only as indicative.

\subsection{Dust formation in other types of cool hydrogen-deficient 
            carbon stars}

Recent analysis of {\it IUE\,} spectra of other hydrogen-deficient 
carbon stars has revealed possible differences between the RCB stars 
and the (optically) spectroscopically-similar HdC stars (Lambert 1986).  
Brunner, Clayton \& Ayres (1998) found no evidence of \ion{C}{2} 
$\lambda$1335 in the HdC star HD 182040, whereas the line is present 
at all times in the UV spectrum of V854 Cen (Fig. 4), R CrB (Holm 
et al. 1987) and RY Sgr (Clayton et al. 1999).  The line may also be 
absent in XX Cam (Brunner et al. 1998), which has been classified
as both RCB and HdC star, but has no bright-IR excess like HD 182040.
Absent or weak \ion{C}{2} $\lambda$1335 may be an important 
discriminator between RCB and HdC stars.  Lawson \& Cottrell (1997)
found that the HdC stars are either low-amplitude pulsators, or that 
they are not pulsating above a 1-$\sigma$ radial velocity limit of 
$\sim$ 1.5 km\,s$^{-1}$.  Only 1 of 5 HdC stars measured by Lawson 
\& Cottrell, HD 175883, was found to have a photometric and radial 
velocity amplitude comparable to the RCB stars.  HD 175893, along 
with HD 173409, was suspected of having a weak (compared to RCB 
stars) infrared excess in {\it IRAS\,} 12-\micron\, photometry 
(Walker 1986).  {\it ISO\,} 12- and 25-\micron\, photometry confirms 
that only the excess in HD 175893 is real (Lawson et al., in 
preparation).  Pulsations in these types of stars seem responsible
for encouraging mass-loss in the form of high-velocity outflows and 
dust, but only if the radial velocity amplitude exceeds 10--15
km\,s$^{-1}$ peak-to-peak.  The dust and gas may be linked by the 
presence of 200--400 km\,s$^{-1}$ blue-shifted absorption seen during 
the declines (Alexander et al. 1972, Cottrell et al. 1990, Clayton 
et al. 1993).  This may be gas being dragged away from the star by 
the ejected dust cloud.  Gas density enhancement in the BL region 
due to enhanced mass-loss in RCB stars, compared to HdC stars, may 
be responsible for the strong \ion{C}{2} $\lambda$1335 emission.

\subsection{V854 Cen as a pulsating star?}

All well-observed RCB stars are pulsating stars.  Most RCB stars 
with $T_{\rm eff}$ similar to V854 Cen (such as RY Sgr and R CrB;
$T_{\rm eff} \approx$ 7000 K), have a radial velocity-to-light 
(RV/$V$) amplitude ratio of $\approx$ 50 km\,s$^{-1}$\,mag$^{-1}$, 
similar to radially-pulsating Cepheids (Lawson \& Cottrell 1997).  
Typical amplitudes (peak-to-peak) are 10--15 km\,s$^{-1}$ in radial 
velocity and 0.2--0.3 in $V$, although RY Sgr is more active 
(30--40 km\,s$^{-1}$ and 0.5--0.7 mag, respectively).  

So far, it has not been possible to reliably measure pulsation
amplitudes for V854 Cen, due to the extreme nature of the light
curve of the star, and the likely low-amplitude of the pulsations. 
Some observations at maximum in the 1989--1991 light curve (Lawson 
et al. 1992) showed semi-regular variations on timescales of $\sim$ 
40-d and with amplitudes of 0.1--0.2 mag that are probably due to 
radial pulsations.  Lawson \& Cottrell (1989) did not detect radial 
velocity variations in a short series of measurements made in 1988, 
but the individual measurements had 1-$\sigma$ uncertainties of 
3--5 km\,s$^{-1}$.  If RV/$V \approx 50$ for V854 Cen, like other 
RCB stars, then the radial velocity amplitude is expected to be 
only $\sim$ 10 km\,s$^{-1}$.   

The onset times of declines in V854 Cen are satisfied by a 42.23 d
period, which is probably the pulsation period of the star.  Other
RCB stars of similar $T_{\rm eff}$ have similar periods (Lawson
et al. 1990, Lawson \& Cottrell 1997).  It remains unresolved why
V854 Cen is curently more active than other RCB stars.  The greater 
hydrogen abundance in V854 Cen, compared to any other known RCB star, 
may encourage dust production.  However, other RCB stars are known 
to have experienced prolonged intervals of dust production in the 
past.

\acknowledgments

WAL thanks the University College ADFA Special Research Grant 
Scheme, and Department of Physics and Astronomy at LSU for 
financial support.  WAL and MMM thank the Australian Research 
Council Small Grant Scheme FY97 for supporting this research.
GCC was supported by NASA grant JPL 961526.  

{}

\clearpage

\figcaption{Visual and photoelectric $V$ light curve for V854 Cen 
from 1987--1998.  Times of the {\it IUE\,} SWP and LWP spectra are 
indicated.  Also shown are calculated decline onset times from the 
pulsation-decline ephemeris derived in Section 4.}

\figcaption{Color curves for V854 Cen from 1989--1998, including 
decline onset times over this interval.}

\figcaption{The 1991--1994 light curve for V854 Cen on an expanded
scale, with times of the {\it IUE\,} SWP and LWP spectra indicated
(top panel), and integrated line fluxes (line intensities) for key 
emission features in the SWP and LWP spectra (lower panels).  Line 
intensity units are erg\,s$^{-1}$\,cm$^{-2}$.  Some measurements 
could not made due to the intrusion of bright stellar continuum 
(usually when the star was at or near maximum light).  The vertical
bar in the lower-right corner of the panel for each emission line 
indicates a nominal uncertainty of $\pm$ 20 percent in the line 
intensities, based upon the average flux for each line (see
Section 3.3 for further discussion of the uncertainties).}

\figcaption{(upper) Sample SWP spectra of V854 Cen obtained at 
maximum and minimum light. (lower) Average of SWP spectra of 
V854 Cen obtained at minimum, constructed by co-adding the 5 
long-exposure ($>$ 12 ks) spectra obtained during declines 
that are without significant photospheric absorption longward
of 1500 \AA, weighted by exposure time. Key emission features 
are labelled.} 

\figcaption{Sample LWP spectra of V854 Cen obtained at maximum 
(upper) and minimum light (lower, where the re-scaled bright 
spectrum is also shown for comparison).  Key absorption and 
emission features are labelled.}

\figcaption{Decay of the LWP emission lines during the 1991 
decline.  The lines are exponential fits to the emission line 
intensities.  Decay time-scales (half-lives) for the emission 
lines range from 50--200~d (see Section 3.3.1 and Table 3).
A nominal uncertainty of $\pm$ 20 percent in the average 
line intensity of the LWP lines is indicated in the upper
panel.}

\figcaption{(upper) Observed--Calculated (O--C) residuals 
versus cycle number for the pulsation-decline ephemeris for 
V854 Cen.  (lower) O--C residuals plotted as a histogram.} 


\clearpage

\begin{thebibliography}{}
\bibitem[]{}Alexander, J. B., Andrews, P. J., Catchpole, R. M., Feast, 
            M. W., Lloyd Evans, T., Menzies, J. W., Wisse, P. N. J., \& 
            Wisse, M. 1972, \mnras, 158, 305
\bibitem[]{}Brunner, A. R., Clayton, G. C., \& Ayres, T. R. 1998, 
            \pasp, 110, 1412
\bibitem[]{}Clayton, G. C., Whitney, B. A., Stanford, S. A., Drilling, 
            J. S.,  \& Judge, P. G. 1992a, \apjl, 384, L19
\bibitem[]{}Clayton, G. C., Whitney, B. A., Stanford, S. A., \& Drilling, 
            J. S. 1992b, \aj, 397, 652
\bibitem[]{}Clayton, G. C., Lawson, W. A., Whitney, B. A., \& Pollacco,
            D. L. 1993, \mnras, 264, L13
\bibitem[]{}Clayton, G. C., Lawson, W. A., Cottrell, P. L., Whitney,
            B. A., Stanford, S. A., \& de Ruyter, F. 1994, \apj,
            432, 785
\bibitem[]{}Clayton, G. C., Ayres, T. R., Lawson, W. A., Drilling,
            J. S., Woitke, P., Asplund, M. 1999, \apjl, in press
\bibitem[]{}Cottrell, P. L., Lawson, W. A., \& Buchhorn, M. 1990,
            \mnras, 244, 149
\bibitem[]{}Evans, A., Whittet, D. C. B., Davies, J. K., Kilkenny, D.
           , \& Bode M. F. 1985, \mnras, 217, 767 
\bibitem[]{}Feast, M. W. 1986, Hydrogen Deficient Stars and Related
            Objects, K. Hunger, Dordrecht: Reidel, 151
\bibitem[]{}Herbig, G. H. 1949, \apj, 110, 143
\bibitem[]{}Holm, A. V., Hecht, J., Wu, C. -C., \& Donn, B. 1987, \pasp,
            99, 497
\bibitem[]{}Holm, A. V., \& Wu, C. -C. 1982, Advances in Ultraviolet
            Astronomy: Four Years of {\it IUE\,} Research, Y. Kondo, 
            J. Mead and R. C. Chapman, Washington, DC: NASA, 429
\bibitem[]{}Jordan, C., \& Linsky, J. L. 1987, Exploring the Universe
            with the {\it IUE\,} satellite, Y. Kondo, Dordrecht: Kluwer,
            259 
\bibitem[]{}Kilkenny, D. 1982, \mnras, 200, 1019
\bibitem[]{}Kilkenny, D., \& Marang, F. 1989, \mnras, 238, 1P
\bibitem[]{}Lambert, D. L. 1986, Hydrogen Deficient Stars and Related
            Objects, K. Hunger, Dordrecht: Reidel, 127
\bibitem[]{}Lawson, W. A. 1992, \mnras, 258, 1P
\bibitem[]{}Lawson, W. A., \& Cottrell, P. L. 1989, \mnras, 240, 689
\bibitem[]{}Lawson, W. A., \& Cottrell, P. L. 1997, \mnras, 285, 266
\bibitem[]{}Lawson, W. A., Cottrell, P. L., Gilmore, A. C., \& 
            Kilmartin, P.M. 1990, \mnras, 247, 91
\bibitem[]{}Lawson, W. A., Cottrell, P. L., \& Clark, M. 1991, 
            \mnras, 251, 687
\bibitem[]{}Lawson, W. A., Cottrell, P. L., Gilmore, A. C., \& 
            Kilmartin, P.M. 1992, \mnras, 256, 339
\bibitem[]{}Marraco, H. G., \& Milesi, G. E. 1982, \aj, 87, 1775 
\bibitem[]{}McNaught, R., \& Dawes, G. 1986, IAU Circ. 4233
\bibitem[]{}Payne-Gaposchkin, C. 1963, \apj, 138, 320
\bibitem[]{}Pugach, A. F. 1977, Info. Bull. var. Stars, No. 1277
\bibitem[]{}Pugach, A. F. 1990, AZh, 67, 1280
\bibitem[]{}Rao, N. K., Nandy, K., \& Bappu, M. K. V. 1981, \mnras,
            195, 71P
\bibitem[]{}Schartel, N., \& Skillen, I. 1998, UV Astrophysics,
            Beyond the IUE Final Archive, W. Wamsteller and R. Gonzalez
            Riestra, ESTEC: Noordwijk, 735
\bibitem[]{}Sterken, C., \& Jones, A. F. 1997, J. Astron. Data,
            3, 4 
\bibitem[]{}Walker, H. J. 1986, Hydrogen Deficient Stars and Related
            Objects, K. Hunger, Dordrecht: Reidel, 407
\bibitem[]{}Whitney, B. A., Clayton, G. C., Schulte-Ladbeck, R. E. 
           , \& Meade, M. R. 1992, \aj, 103, 1652
\bibitem{}{}Woitke, P., Goeres, A., \& Sedlmayr, E. 1996, \aap, 313, 217
\bibitem[]{}Wu, C. -C., Reichert, G. A., Ake, T. B., Boggess, A.,
            Holm, A. V., Imhoff, C. L., Kondo, Y., Mead, J. M.
           , \& Shore, S. N. 1982, International Ultraviolet
            Explorer ({\it IUE\,}) Ultraviolet Spectral Atlas of Selected
            Astronomical Objects, Washington, DC: NASA, 4
\end{thebibliography}
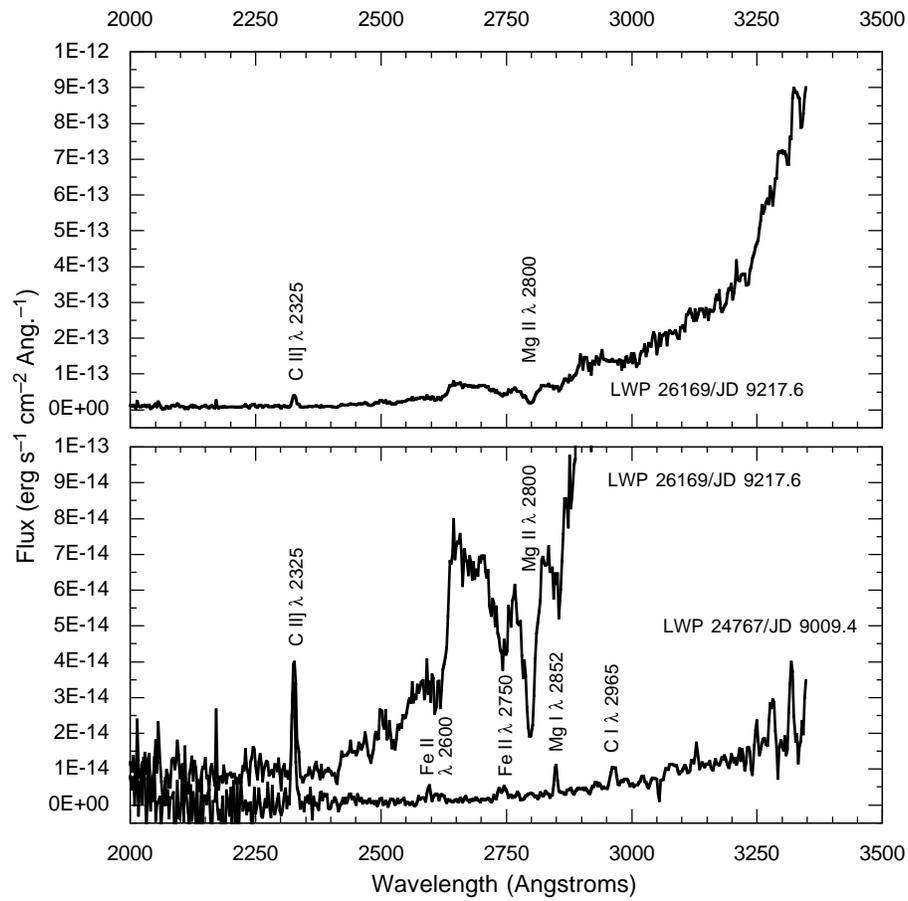
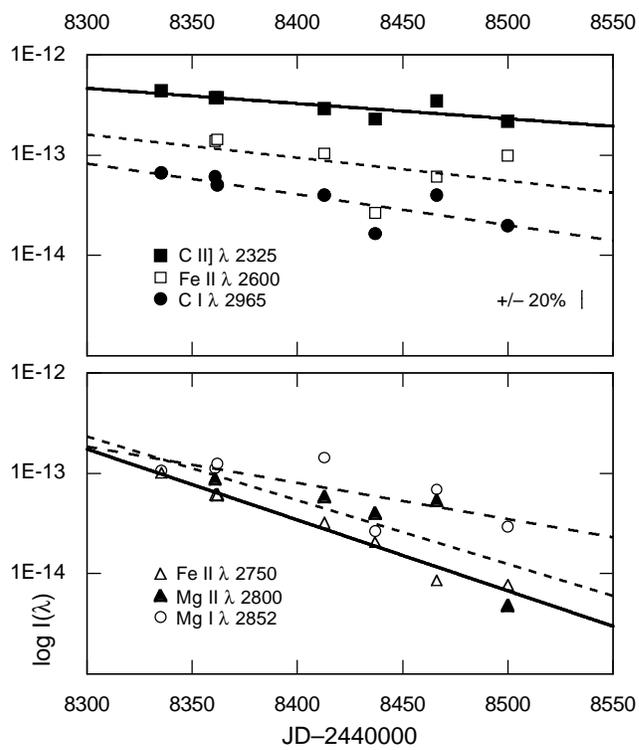
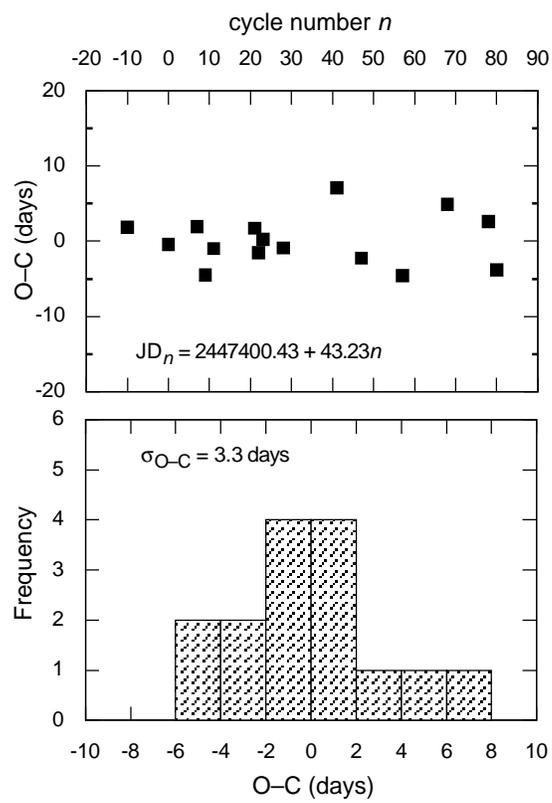
\end{document}